\documentclass[preprint,prc,aps,showpacs,groupedaddress,superscriptaddress,floatfix]{revtex4}
\usepackage{amsfonts}
\usepackage{amsmath}

\input{tcilatex}

\begin{document}

\title{Rotational and vibrational diatomic molecule in the Klein-Gordon
equation with hyperbolic scalar and vector potentials\textbf{\ }}
\author{Sameer M. Ikhdair}
\email[E-mail: ]{sameer@neu.edu.tr; sikhdair@neu.edu.tr}
\affiliation{Department of Physics, Near East University, Nicosia, Cyprus, Turkey}
\date{\today }

\begin{abstract}
We present an approximate analytic solution of the Klein-Gordon equation in
the presence of equal scalar and vector generalized deformed hyperbolic
potential functions by means of parameteric generalization of the
Nikiforov-Uvarov method. We obtain the approximate bound state
rotational-vibrational (ro-vibrational) energy levels and the corresponding
normalized wave functions expressed in terms of the Jacobi polynomial $%
P_{n}^{\left( \mu ,\nu \right) }(x),$ where $\mu >-1,$ $\nu >-1$ and $x\in %
\left[ -1,+1\right] $ for a spin-zero particle in a closed form. Special
cases are studied including the non-relativistic solutions obtained by
appropriate choice of parameters and also the $s$-wave solutions.

\end{abstract}

\maketitle

\bigskip

\section{Introduction}

\noindent In nuclear, molecular and high energy physics [1,2], one of the
interesting problems is to obtain exact solutions of the relativistic wave
equations like Klein-Gordon (KG), Dirac and Salpeter wave equations for
mixed vector and scalar potential. The KG equation has also been used to
understand the motion of a spin-zero particle in large class of potentials
using different methods. This allows us to introduce two types of potential
coupling, which are the four-vector potential ($V$) and the space-time
scalar potential ($S$).

For the case $S=\pm V,$ the solution of these wave equations with physical
potentials has been studied recently. Furthermore, their exact solutions are
possible only for certain central potentials such as Morse potential [3],
Hulth\'{e}n potential [4], Woods-Saxon potential [5], P\"{o}schl-Teller
potential [6], reflectionless-type potential [7], pseudoharmonic oscillator
[8], ring-shaped harmonic oscillator [9], $V_{0}\tanh ^{2}(r/r_{0})$
potential [10], five-parameter exponential potential [11], Rosen-Morse
potential [12], and generalized symmetrical double-well potential [13], etc
by using different methods. It is remarkable that in most of these works,
the scalar and vector potentials are almost taken to be equal (i.e., $S=V$)
[2,14]. Recently, interest in the solutions of the KG equation for the case
where $S(r)=\pm V(r)$ has surged. It presents bounded solutions in the
relativistic view, although the KG equation reduces to the Schr\"{o}%
dinger-like problem in the nonrelativistic limit. However, the reduced
equation can show the relativistic properties of the system. For the most
recent contributions, one may consult the papers in [2,6,9,10] and
references therein. Therefore, the choice $S(r)=V(r)$ (positive energy
states) produces a nontrivial nonrelativistic limit with a potential
function $\Sigma (r)=2V(r),$ and not $V(r).$ It represents the concept of
the exact spin symmetry that occurs in nuclei (\textit{i.e.}, when the
difference potential $\Delta (r)=V(r)-S(r)=0).$ In the negative energy
states (corresponding to $S(r)=-V(r)$) the nonrelativistic limit is the
trivial interaction free-mode. It represents the concept of the pseudospin
symmetry (\textit{i.e.}, occurs when the sum potential $\Sigma
(r)=V(r)+S(r)=0)$ [15]. The spin symmetry is relevant for mesons [16]. The
pseudospin symmetry concept has been applied to many systems in nuclear
physics and related areas [15-17] and used to explain features of deformed
nuclei [18], the super-deformation [19] and to establish an effective
nuclear shell-model scheme [17,20]. This, of course, does not diminish the
importance of such problems. It only limits it's contribution (with the
proper physical interpretation) to the relativistic regime [14]. However, in
some few other cases, it is considered the case where the scalar potential
is greater than the vector potential (in order to guarantee the existence of
KG bound states) (i.e., $S>V$) [21-24]. Many authors have considered a more
general transformation between the unequal vector and scalar potentials
given by $V(r)=V_{0}+\beta S(r)$ (or $S(r)\neq \pm V(r)$)$,$where $V_{0}$
and $\beta $ being arbitrary constants of certain proportions have to be
chosen after solving the problem under consideration (cf. Ref. [25] and
references therein). Nonetheless, such physical potentials are very few. The
bound state solutions for the last case is being obtained for the
exponential potential with the $s$-wave ($l=0$) KG equation when the scalar
potential is greater than the vector potential [21].

The problems connected with the molecular structure provide interesting and
instructive applications of quantum mechanics, since molecules are
considerably more complex in structure than atoms. Two distinct problems
arise in connection with molecular structure. The first is to obtain the
electronic wave functions and potential energy functions of the nuclear
coordinates. This problem can be solved analytically only in the simplest
cases. The second is to obtain the solution of the nuclear motion equation.
In this regard, the construction of a suitable potential function of a
diatomic molecule is very important. It has been found that the
potential-energy function for the lowest electronic states of actual
diatomic molecules can be expressed by means of the Morse potential [26]:%
\begin{equation}
V_{M}(r)=D\left[ 1-\exp \left[ -\alpha (r-r_{e})\right] \right] ^{2},
\end{equation}%
which has three adjustable positive parameters $\alpha ,$ $D$ and $r_{e}.$
At $r=r_{e},$ it has a minimum value at zero and approaches $D$
exponentially for large $r.$ If $\frac{1}{\alpha }$ is somewhat smaller than 
$r_{e},$ it becomes large (but not infinie) as $r\rightarrow 0.$ This
potential is important in the field of molecular physics describing the
interaction between two atoms [26,27]. Overmore, progress has been made in
the field of diatomic molecules and extensive use of the potential functions
have been introduced [28,29]. At present the Morse potential is still one of
the potential functions used most in molecular physics and quantum chemistry
[30]. However, it has few asymptotic inaccuracies in the regions of small
and large $r.$ To avoid these inaccuracies, many works have been carried out
in that direction to improve Morse potential [31]. In 1986, Schi\"{o}berg
[32] suggested hyperbolical (empirical) potential functions of the form: 
\begin{equation}
V_{\pm }(r)=D\left\{ 1-\sigma \left[ \coth (\alpha r)\right] ^{\pm
1}\right\} ^{2},
\end{equation}%
where $D,$ $\alpha $ and $\sigma $ are three adjustable positive parameters
with $D=D_{e}/(1-\sigma )^{2}$ ($D_{e}$ is the spectroscopic dissociation
energy) and $\sigma \rightarrow \sigma /\delta ,$ $\delta \neq 0$ is a
parameter. In contrast to the everywhere-regular Morse potential, $V_{+}(r)$
is highly singular at the origin with $1/r^{2}$ and $1/r$ singularities. The
two potentials behave similarly near the extremum point $r=r_{e}.$ It has
the minimum value $0$ at the point%
\begin{equation}
r=r_{e}=\frac{1}{\alpha }\arctan h\left( \sigma \right) ^{\pm 1},
\end{equation}%
and approaches $D$ exponentially for large $r.$ Unlike the Morse potential
(1), the empirical potential (EP) function $V_{+}(r)$ approaches infinity at
the point $r=0.$ In the region of large $r,$ it is closer to the
experimental Rydberg-Klein-Rees (RKR) curve than the Morse potential for
some diatomic molecules. For a diatomic molecular model instead of the
nuclear model, we consider the reduced mass definition. If the nuclei have
masses $m_{1}$ and $m_{2},$ the reduced mass is defined as $\mu
=m_{1}m_{2}/(m_{1}+m_{2})$ and at this point the diatomic molecular model
can be included to the the pseudospin symmetry concept. In this context, it
is worth noting that the energy of the nuclear motion (in $MeV$) is widely
separated from the energy in atomic vibration and rotation (in $KeV$). That
is why, one is able to separate the two motions and study the atomic
vibration and rotation separately from the nuclear motion.

Lu \textit{et al} [33] obtained an approximate solution for the Schr\"{o}%
dinger equation of diatomic molecule oscillator with the positive sign EP
functions given in Eq. (2) ($\delta =1$ case) by means of the hypergeometric
series method. Further, rigorous energy eigenvalues and eigenfunctions for
the $1D$ Schr\"{o}dinger eqauation are also obtained using a similar method$%
. $ Since there are no exact analytical solutions for the EP functions $%
(l\neq 0)$ without an approximation to the centrifugal term $\sim 1/r^{2},$
some approximation [34] was used to obtain these solutions. One of these
approximations have been employed in solving the rotating Morse potential
for any $l$-state [34,35]. The ro-vibrating energy eigenvalues of the EP
functions were determined with a semiclassical (SC) procedure (the
Bohr-Sommerfeld quantization condition) and a quantum-mechanical (QM) method
(the Schr\"{o}dinger equation) [33]. Furthermore, several approximation
schemes have been developed to find better analytical formulas for Eq. (2)
with $\delta =1$ [33,36]. Overmore, Jia \textit{et al} [37] have used the
basic concepts of the supersymmetric quantum mechanics formalism and the
functional analysis method to investigate approximately the pseudospin
symmetric solutions of the Dirac equation for the arbitrary pseudo-orbital
angular momentum number $\widetilde{l}$ and to obtain the bound state
solutions for the nuclei in the relativistic EP as a diatomic molecular
model. Very recently, the NU method [38-40] was applied to solve the radial
Schr\"{o}dinger wave equation with the EP functions for $l\neq 0$ case [41].
The analytic solution is used to obtain the ro-vibrating energy states for
selected $H_{2}$ and $Ar_{2}$ diatomic molecules using the relevant
potential parameters and spectroscpic constants given in Ref. [32].

Over the past years, the quantum deformation [42] has been the subject of
interest because of its relevance with applications in nuclei [43-45],
statistical quantum theory, string beam theory and conformal field theory
[46-49]. Recently, some authors have introduced few potentials in terms of
hyperbolic functions [50,51] in the view of $q$-deformation [52].

Encouraged with the high performance of the above inter-molecular potential,
we write the EP functions as 
\begin{subequations}
\begin{equation}
\left[ \coth \alpha (r-r_{e})\right] ^{\pm 1}=\frac{e^{\alpha (r-r_{e})}\pm
e^{-\alpha (r-r_{e})}}{e^{\alpha (r-r_{e})}\mp e^{-\alpha (r-r_{e})}}=\frac{%
1\pm e^{-2\alpha (r-r_{e})}}{1\mp e^{-2\alpha (r-r_{e})}}=\frac{1\pm
qe^{-2\alpha r}}{1\mp qe^{-2\alpha r}}=\left[ \coth _{q}(\alpha r)\right]
^{\pm 1},
\end{equation}%
\begin{equation}
V_{\pm }(r,q)=D\left\{ 1-\sigma \left[ \coth \alpha (r-r_{e})\right] ^{\pm
1}\right\} ^{2}=D\left\{ 1-\sigma \left[ \frac{1\pm qe^{-2\alpha r}}{1\mp
qe^{-2\alpha r}}\right] \right\} ^{2},
\end{equation}%
where $q=e^{2\alpha r_{e}},$ giving a magnitude for $q$ that is larger than
one. For an inverse transformation $q=e^{-2\alpha r_{e}},$ the magnitude for 
$q$ varies between zero and one. In the context of the quantum deformation
[52], the above form is similar to a $q$-deformed (perturbed) generalized
deformed empirical potential ($q$-DEP/GDEP) functions. It is worth noting
that the range of parameter $q$ was taken as $q>0$ in [50] and has been
extended to $-1\leq q<0$ or $q>0$\ or even complex by Ref. [51]. Such $q$%
-deformed potential functions have been introduced for the first time by
Arai [42] for real $q$ values. When\ $q$ is complex, these functions are
called the generalized deformed potential functions. In this paper, we
intend to find the analytic solution of the KG equation for the equal scalar
and vector $q$-DEP/GDEP with any orbital angular quantum number ($l\neq 0).$
The specific choice of $V(r)=S(r)$ allows one to make KG equation
approximately soluble for it's relativistic energy eigenvalues and wave
functions. Further, it opens up a new approach of generating the
non-relativistic solution which is found to coincide with the previous Schr%
\"{o}dinger solution of Eq. (4). In the present calculations, we apply a
parameteric generalization procedures of the NU method which are making our
calculations straightforward and simple.

The present paper is organized as follows. In sect. 2, we present a
parameteric generalization of the NU method which holds for the
exponential-type potentials. In sect. 3, we obtain an approximate analytic
NU bound state solution of the ($3+1$)-dimensional KG equation for equal
scalar and vector $q$-DEP/GDEP functions with arbitrary $l$-states. In sect.
4, we discuss two special cases, the vibrational ($l=0$) and the
non-relativistic limit (Schr\"{o}dinger solution). In Sect. 5, we calculate
the ro-vibrating energy states for selected $H_{2}$ and $Ar_{2}$ diatomic
molecules in the non-relativistic approach. Section 6 contains the relevant
conclusions.

\section{\noindent NU Method}

The NU method is briefly outlined here and the details can be found in [38].
This method was proposed to solve the second-order differential equation of
the hypergeometric-type: 
\end{subequations}
\begin{equation}
\sigma ^{2}(z)g^{\prime \prime }(z)+\sigma (z)\widetilde{\tau }(z)g^{\prime
}(z)+\widetilde{\sigma }(z)g(z)=0,
\end{equation}%
where $\sigma (z)$ and $\widetilde{\sigma }(z)$ are at most second-degree
polynomials and $\widetilde{\tau }(s)$ is a first-degree polynomial. The
primes denote derivatives with respect to $z.$ To find a particular solution
of Eq. (5), one can decompose the wave functions, $g_{nl}(z)$ as follows:%
\begin{equation}
g(z)=\phi (z)y_{n}(z),
\end{equation}%
leading to recast (5) in the hypergeometric-type equation 
\begin{equation}
\sigma (z)y_{n}^{\prime \prime }(z)+\tau (z)y_{n}^{\prime }(z)+\lambda
y_{n}(z)=0,
\end{equation}%
where%
\begin{equation}
\lambda =k+\pi ^{\prime }(z),
\end{equation}%
and $y_{nl}(z)$ satisfies the Rodrigues relation%
\begin{equation}
y_{n}(z)=\frac{A_{n}}{\rho (z)}\frac{d^{n}}{dz^{n}}\left[ \sigma ^{n}(z)\rho
(z)\right] .
\end{equation}%
In the above equation, $A_{n}$ is a constant related to the normalization
and $\rho (z)$ is the weight function satisfying the condition%
\begin{equation}
\left[ \sigma (z)\rho (z)\right] ^{\prime }=\tau (z)\rho (z),
\end{equation}%
with 
\begin{equation}
\tau (z)=\widetilde{\tau }(z)+2\pi (z),\tau ^{\prime }(z)<0.
\end{equation}%
The weight function should be carefully chosen because it has an influence
on the performance of orthogonal wave functions, orthogonal in the interval $%
[0,1],$ of the type Laguerre $L_{n}^{(\gamma )}(z)$ and Jacobi $%
P_{n}^{\left( \alpha ,\beta \right) }(z)$ polynomials etc. It is defined
with a compact support often called domain of influence which can be spheres
in three-dimensions. Generally speaking, the weight function commonly used
is exponential function. Furthermore, the weighted integral \ and weighted
average are defined by $h(z)=\dint\limits_{\Omega }f(z)\rho (z)dz$ and $%
g(z)=\dint\limits_{\Omega }f(z)\rho (z)dz/\dint\limits_{\Omega }\rho (z)dz,$
if $\ f(z):\Omega \in 
\mathbb{R}
,$ real-valued orthogonal polynomial functions, respectively, with $%
f(z)=\sigma (z)$ and $g(z)=P_{n}^{\left( \alpha ,\beta \right) }(z),$ $%
L_{n}^{(\gamma )}(z),\cdots ,$ etc. Since $\rho (z)>0$ and $\sigma (z)>0,$
the derivative of $\tau (z)$ needs to be negative [38] which is the
essential condition in making the choice of particular solution relevant to
the real bound state solution. The other part of the wave functions in Eq.
(6) is mainly the solution of the logarithmic derivative:%
\begin{equation}
\frac{\phi ^{\prime }(z)}{\phi (z)}=\frac{\pi (z)}{\sigma (z)},
\end{equation}%
where%
\begin{equation}
\pi (z)=\frac{1}{2}\left[ \sigma ^{\prime }(z)-\widetilde{\tau }(z)\right]
\pm \sqrt{\frac{1}{4}\left[ \sigma ^{\prime }(z)-\widetilde{\tau }(z)\right]
^{2}-\widetilde{\sigma }(z)+k\sigma (z)}.
\end{equation}%
is a polynomial of order one. The determination of $k$ is the essential
point in the calculation of $\pi (z),$ for which the discriminant of the
square root in the last equation is set to zero. This gives the polynomial $%
\pi (z)$ which is dependent on the transformation function $z(r).$ Also, the
parameter $\lambda $ defined in Eq. (8) takes the form%
\begin{equation}
\lambda =\lambda _{n}=-n\tau ^{\prime }(z)-\frac{1}{2}n\left( n-1\right)
\sigma ^{\prime \prime }(z),\ \ \ n=0,1,2,\cdots .
\end{equation}%
In this regard, we can construct a parameteric generalization of the NU
method valid for any central and non-central exponential-type potentials. We
begin by comparing the following general hypergeometric equation 
\begin{equation}
\left[ z\left( 1-c_{3}z\right) \right] ^{2}g^{\prime \prime }(z)+\left[
z\left( 1-c_{3}z\right) \left( c_{1}-c_{2}z\right) \right] g^{\prime
}(z)+\left( -B_{1}z^{2}+B_{2}z-B_{3}\right) g(z)=0,
\end{equation}%
with it's counterpart equation (5) to obtain%
\begin{equation*}
\widetilde{\tau }(z)=c_{1}-c_{2}z,
\end{equation*}%
\begin{equation*}
\sigma (z)=z\left( 1-c_{3}z\right) ,
\end{equation*}%
\begin{equation}
\widetilde{\sigma }(z)=-B_{1}z^{2}+B_{2}z-B_{3}.
\end{equation}%
where the parameters $c_{i}$ and $B_{i}$ ($i=1,2,3$) are constants to be
determined during the solution process. Thus, following the method, we may
obtain all the analytic polynomials and their relevant constants necessary
for the solution of a radial wave equation. These analytic expressions are
given in Appendix A.

\section{The KG Solution of Equal Scalar-Vector GDEP Functions}

In relativistic quantum mechanics, we usually use the KG equation for
describing a scalar particle, i.e., the spin-$0$ particle dynamics. The
discussion of the relativistic behavior of spin-zero particles requires
understanding the single particle spectrum and the exact solutions to the KG
equation which are constructed by using the four-vector potential $\mathbf{A}%
_{\lambda }$ $(\lambda =0,1,2,3)$ and the scalar potential $(S)$. In order
to simplify the analytic solution of the KG equation, the four-vector
potential can be written as $\mathbf{A}_{\lambda }=(A_{0},0,0,0).$ The first
component of the four-vector potential is represented by a vector potential $%
(V),$ i.e., $A_{0}=V.$ In this case, the motion of a relativistic spin-$0$
particle in a potential is described by the KG equation with the potentials $%
V$ and $S$ [1]$.$ For $S=V$ case [39]$,$ the ($3+1$)-dimensional KG equation
is recasted to a Schr\"{o}dinger-like equation and thereby the bound state
solutions are easily obtained by using the well-known methods developed in
nonrelativistic quantum mechanics [2].

Let us now consider the ($3+1$)-dimensional time-independent KG equation
describing a scalar particle (spin-$0$ particle) with Lorentz scalar $S(r)$
and Lorentz vector $V(r)$ potentials which takes the form [2,14,53]

\begin{equation}
\left[ \mathbf{P}_{op}^{2}-\left( V(r)-E_{R}\right) ^{2}+\left(
S(r)+mc^{2}\right) ^{2}\right] \psi _{KG}(\overrightarrow{r})=0,
\end{equation}%
where $m$ $\ $and $E_{R}$ denote the reduced mass and relativistic binding
energy of two interacting particles, respectively, with $\mathbf{P}%
_{op}=-i\hbar \overrightarrow{\nabla }$ is the momentum operator. It would
be natural to scale the potential terms in Eq. (17) so that in the
nonrelativistic limit the interaction potential becomes $V(r),$ not $2V(r).$
We follow Alhaidari \textit{et al }[14] to reduce the above equation to the
form [54] 
\begin{equation}
\left\{ \mathbf{\nabla }^{2}+\frac{1}{\hbar ^{2}c^{2}}\left[ \left( \frac{1}{%
2}V(r)-E_{R}\right) ^{2}-\left( \frac{1}{2}S(r)+mc^{2}\right) ^{2}\right]
\right\} \psi _{KG}(\overrightarrow{r})=0.
\end{equation}%
Thus, after making use of the equal scalar and vector GDEP functions ($%
S_{\pm }(r;q)=V_{\pm }(r;q)),$ Eq. (18) recasts to 
\begin{equation*}
\left\{ \mathbf{\nabla }^{2}-\frac{1}{\hbar ^{2}c^{2}}\left[ \alpha
_{2}^{2}\left( \alpha _{1}^{2}+V_{\pm }(r,q)\right) \right] \right\} \psi
_{KG}(\overrightarrow{r})=0,
\end{equation*}%
\begin{equation}
\mathbf{\nabla }^{2}=\frac{\partial ^{2}}{\partial r^{2}}+\frac{2}{r}\frac{%
\partial }{\partial r}+\frac{1}{r^{2}}\left[ \frac{1}{\sin \theta }\frac{%
\partial }{\partial \theta }\left( \sin \theta \frac{\partial }{\partial
\theta }\right) +\frac{1}{\sin ^{2}\theta }\frac{\partial ^{2}}{\partial
\varphi ^{2}}\right] ,\text{ }r^{2}=\text{ }\sum\limits_{j=1}^{3}x_{j}^{2},
\end{equation}%
where $\alpha _{1}^{2}=mc^{2}-E_{R},$ $\alpha _{2}^{2}=mc^{2}+E_{R}.$ It is
woth noting that the solution of the ($3+1$)-dimensional KG equation can be
reduced to the solution of the Schr\"{o}dinger equation with the following
appropriate choice of parameters: $\alpha _{1}^{2}\rightarrow -E_{NR\text{ }%
} $ and $\alpha _{2}^{2}/\hbar ^{2}c^{2}\rightarrow 2\mu /\hbar ^{2},$ where 
$\mu =m_{1}m_{2}/(m_{1}+m_{2})$ is the reduced atomic mass for the diatomic
molecular system [14,54]$.$ In addition, we take the interaction potential
as in (4) and decompose the total wave function $\psi _{KG}(\overrightarrow{r%
}),$ with a given angular momentum $l,$ as a product of a radial wave
function $R_{l}(r)=\frac{g(r)}{r}$ and the angular dependent spherical
harmonic functions $Y_{lm}(\widehat{r})$: [53-56] 
\begin{equation}
\psi _{KG}(\overrightarrow{r})=\frac{g(r)}{r}Y_{lm}(\widehat{r}),
\end{equation}%
with angular momentum quantum numbers being $l$ and $m.$ This reduces Eq.
(19) into the form%
\begin{equation}
\frac{d^{2}g(r)}{dr^{2}}-\frac{1}{\hbar ^{2}c^{2}}\left[ \alpha
_{1}^{2}\alpha _{2}^{2}+\alpha _{2}^{2}D\left[ 1-\sigma \left( \frac{1\pm
q\exp (-2\alpha r)}{1\mp q\exp (-2\alpha r)}\right) \right] ^{2}+\frac{%
l(l+1)\hbar ^{2}c^{2}}{r^{2}}\right] g(r)=0,\text{ }q\neq 0,
\end{equation}%
where $\frac{l(l+1)}{r^{2}}$ is the centrifugal potential and the boundary
conditions $g(0)=g(\infty )=0$ as we are dealing with bound-state solutions$%
. $ Moreover, if $l$ is not too large, the case of the vibrations of small
amplitude about the minimum, we can then use the approximate expansion of
the centrifugal potential near the minimum point $r=r_{e}$ as [33]%
\begin{equation}
\frac{l(l+1)}{r^{2}}\approx \frac{l(l+1)}{r_{e}^{2}}\left\{ A_{0}+A_{1}\frac{%
\pm \exp (-2\alpha r)}{1\mp q\exp (-2\alpha r)}+A_{2}\left[ \frac{\pm \exp
(-2\alpha r)}{1\mp q\exp (-2\alpha r)}\right] ^{2}\right\} ,
\end{equation}%
where 
\begin{subequations}
\begin{equation}
A_{0}=1-\left[ \frac{1\mp \exp (-2\alpha r_{e})}{2\alpha r_{e}}\right] ^{2}%
\left[ \frac{8\alpha r_{e}}{1\mp \exp (-2\alpha r_{e})}-3-2\alpha r_{e}%
\right] ,
\end{equation}%
\begin{equation}
A_{1}=\pm 2\left[ \exp (2\alpha r_{e})\mp 1\right] \left\{ 3\left[ \frac{%
1\mp \exp (-2\alpha r_{e})}{2\alpha r_{e}}\right] -\left( 3+2\alpha
r_{e}\right) \left[ \frac{1\mp \exp (-2\alpha r_{e})}{2\alpha r_{e}}\right]
\right\} ,
\end{equation}%
\begin{equation}
A_{2}=\left[ \exp (2\alpha r_{e})\mp 1\right] ^{2}\left[ \frac{1\mp \exp
(-2\alpha r_{e})}{2\alpha r_{e}}\right] ^{2}\left[ 3+2\alpha r_{e}-\frac{%
4\alpha r_{e}}{1\mp \exp (-2\alpha r_{e})}\right] ,
\end{equation}%
and higher-order terms are neglected. In fact, Eq. (22) is the approximate
expansion of the centrifugal potential $\frac{l(l+1)}{r^{2}}$ and is valid
for all $r\approx r_{e}$, the minimum point of $V_{\pm }(r)$ since $r$ is
not singular there$.$ However, the expansion is not valid near the
singularity point $r=0.$ Overmore, it is a good approximation for small
vibrations around the equilibrium separation $r-r_{e}.$ When $l\neq 0,$ we
have to use an approximation for the centrifugal term similar to the
non-relativistic cases which is valid only for $q=1$ value [33,53]. However,
for $s$-waves, we remark that the problem can be solved exactly and the
solution is valid for any deformation parameter $q.$ At this point, it is
important to mention that very similar expressions to the above expression
for the energy states have also been found over the past years for the
hyperbolical (exponential-type) potentials with $\delta \rightarrow 1$ in
Eq. (2) for $V_{+}(r)$ (cf. Ikhdair and Sever in Ref. [36]). Very recently,
a new improved approximation scheme [36,39] for the centrifugal potential
term $l(l+1)/r^{2}$ was proposed which appears to be very different from the
ones used by Refs. [33,53,54].

Putting $z=\pm \exp (-2\alpha r)\in $ $\left( \pm 1,0\right) $ for $V_{\pm
}(r),$ and defining the parameters 
\end{subequations}
\begin{subequations}
\begin{equation}
B_{1}=q^{2}\widetilde{K}_{nl}^{2}+\widetilde{S}_{l}^{2}-q\widetilde{Q}_{l}-%
\frac{q^{2}}{4},\text{ }B_{2}=2q\widetilde{K}_{nl}^{2}-\widetilde{Q}_{l},%
\text{ }B_{3}=\widetilde{K}_{nl}^{2},
\end{equation}%
with 
\end{subequations}
\begin{subequations}
\begin{equation}
\widetilde{K}_{nl}=\frac{1}{2\alpha \hbar c}\sqrt{\alpha _{2}^{2}D\left(
1-\sigma \right) ^{2}+\frac{l(l+1)\hbar ^{2}c^{2}}{r_{e}^{2}}A_{0}+\alpha
_{1}^{2}\alpha _{2}^{2}}>0,
\end{equation}%
\begin{equation}
\widetilde{Q}_{l}=-\frac{q\alpha _{2}^{2}D}{\alpha ^{2}\hbar ^{2}c^{2}}%
\sigma \left( 1-\sigma \right) +\frac{l(l+1)}{4\alpha ^{2}r_{e}^{2}}A_{1},
\end{equation}%
\begin{equation}
\widetilde{S}_{l}=\frac{1}{2\alpha \hbar c}\sqrt{4q^{2}\alpha
_{2}^{2}D\sigma ^{2}+\frac{l(l+1)\hbar ^{2}c^{2}}{r_{e}^{2}}%
A_{2}+q^{2}\alpha ^{2}\hbar ^{2}c^{2}}>0.
\end{equation}%
we obtain the hypergeometric wave equation 
\end{subequations}
\begin{equation}
g^{\prime \prime }(z)+\frac{\left( 1-qz\right) }{z\left( 1-qz\right) }%
g^{\prime }(z)+\frac{1}{z^{2}\left( 1-qz\right) ^{2}}\left\{
-B_{1}z^{2}+B_{2}z-B_{3}\right\} g(z)=0,
\end{equation}%
where $g(z)=g(r).$ If we apply the previous transformations, the above
expressions reduce into their non-relativistic limits:

\begin{subequations}
\begin{equation}
\widetilde{K}_{nl}\rightarrow K_{nl}=\frac{1}{2\alpha \hbar }\sqrt{2\mu
D\left( 1-\sigma \right) ^{2}+\frac{l(l+1)\hbar ^{2}}{r_{e}^{2}}A_{0}-2\mu
E_{NR}}>0,
\end{equation}%
\begin{equation}
\widetilde{Q}_{l}\rightarrow Q_{l}=-\frac{2\mu qD}{\alpha ^{2}\hbar ^{2}}%
\sigma \left( 1-\sigma \right) +\frac{l(l+1)}{4\alpha ^{2}r_{e}^{2}}A_{1},
\end{equation}%
\begin{equation}
\widetilde{S}_{l}\rightarrow S_{l}=\frac{q}{2\alpha \hbar }\sqrt{8\mu
D\sigma ^{2}+\frac{l(l+1)\hbar ^{2}}{r_{e}^{2}}\frac{A_{2}}{q^{2}}+\alpha
^{2}\hbar ^{2}}>0,
\end{equation}%
and also when the deformation parameter $q=1,$ the above equations reduce to
their counterparts as in Refs. [33,41].

Now comparing Eq. (26) with Eq. (5), we obtain particular values for the set
of constant parameters given in Section 2: 
\end{subequations}
\begin{equation*}
c_{1}=1,\text{ }c_{2}=c_{3}=q,\text{ c}_{4}=0,\text{ }c_{5}=-\frac{q}{2},%
\text{ }c_{6}=q^{2}\widetilde{K}_{nl}^{2}+\widetilde{S}_{l}^{2}-q\widetilde{Q%
}_{l},
\end{equation*}%
\begin{equation*}
c_{7}=-2q\widetilde{K}_{nl}^{2}+\widetilde{Q}_{l},\text{ }c_{8}=\widetilde{K}%
_{nl}^{2},\text{ }c_{9}=\widetilde{S}_{l}^{2},
\end{equation*}%
\begin{equation*}
c_{10}=2\widetilde{K}_{nl}=2c_{12}>-1,\text{ }c_{11}=\frac{2}{q}\widetilde{S}%
_{l}=2c_{13}-1>-1,
\end{equation*}%
\begin{equation}
c_{12}=\widetilde{K}_{nl}>0,\text{ }c_{13}=\frac{1}{q}\left( \widetilde{S}%
_{l}+\frac{q}{2}\right) >0.
\end{equation}%
Using Eqs. (28) together with Appendix A, we find the following particular
physical solutions for the parameters:%
\begin{equation}
\pi (z)=\widetilde{K}_{nl}-\left( \frac{q}{2}+q\widetilde{K}_{nl}+\widetilde{%
S}_{l}\right) z,
\end{equation}%
\begin{equation}
k=-\widetilde{Q}_{l}-2\widetilde{K}_{nl}\widetilde{S}_{l},
\end{equation}%
and%
\begin{equation}
\tau (z)=1+2\widetilde{K}_{nl}-2\left( q+q\widetilde{K}_{nl}+\widetilde{S}%
_{l}\right) z,
\end{equation}%
where $\tau ^{\prime }(z)=\frac{d\tau (z)}{dz}=-2\left( q+q\widetilde{K}%
_{nl}+\widetilde{S}_{l}\right) <0$ which gives possible real solutions. In
what follows, from Appendix A, we find the ro-vibrational energy equation
with the aid of (28) as

\begin{equation}
2\widetilde{K}_{nl}=\frac{\left( \frac{\widetilde{S}_{l}}{q}\right) ^{2}-%
\frac{\widetilde{Q}_{l}}{q}-\frac{1}{4}-\left( \frac{\widetilde{S}_{l}}{q}+n+%
\frac{1}{2}\right) ^{2}}{\frac{\widetilde{S}_{l}}{q}+n+\frac{1}{2}},\text{ }%
q\neq 0,
\end{equation}%
which can be written more explicitly as%
\begin{equation*}
2\sqrt{\left( mc^{2}+E_{R}\right) D\left( 1-\sigma \right) ^{2}+\frac{%
l(l+1)\hbar ^{2}c^{2}}{r_{e}^{2}}A_{0}+m^{2}c^{4}-E_{R}^{2}}=
\end{equation*}%
\begin{subequations}
\begin{equation}
\frac{4\left( mc^{2}+E_{R}\right) D\sigma +\frac{l(l+1)\hbar ^{2}c^{2}}{%
r_{e}^{2}}\left( \frac{A_{2}}{q^{2}}-\frac{A_{1}}{q}\right) -\left( 
\widetilde{E}+\alpha \hbar c\left( 2n+1\right) \right) ^{2}}{\widetilde{E}%
+\alpha \hbar c\left( 2n+1\right) },
\end{equation}%
\begin{equation}
\widetilde{E}=\sqrt{4\left( mc^{2}+E_{R}\right) D\sigma ^{2}+\frac{%
l(l+1)\hbar ^{2}c^{2}}{r_{e}^{2}}\frac{A_{2}}{q^{2}}+\alpha ^{2}\hbar
^{2}c^{2}},
\end{equation}%
where $n=0,1,2,\cdots $ and $l=0,1,2,\cdots $ signify the usual vibrational
and rotational angular momentum quantum numbers, respectively.

Let us now turn to the calculations of the corresponding wave functions for
the potential under consideration. Thus, referring to the general model in
Appendix A, the explicit form of the weight function reads 
\end{subequations}
\begin{equation}
\rho (z)=z^{2\widetilde{K}_{nl}}(1-qz)^{\frac{2}{q}\widetilde{S}_{l}},
\end{equation}%
which gives the first part of the wave functions (6) as%
\begin{equation}
y_{n}(z)\rightarrow P_{n}^{(2\widetilde{K}_{nl},\frac{2}{q}\widetilde{S}%
_{l})}(1\mp 2qz),\text{ }\widetilde{K}_{nl}>0,\text{ }\widetilde{S}_{l}>0,
\end{equation}%
with the essential requirement that $2\widetilde{K}_{nl}>-1$ and $\frac{2}{q}%
\widetilde{S}_{l}>-1.$ For example, if $q>0$ then $\widetilde{S}_{l}>-\frac{q%
}{2}$ and if $q<0$ then $0<\widetilde{S}_{l}<-\frac{q}{2}.$ Also, the second
part can be found as%
\begin{equation}
\phi ^{\pm }(z)\rightarrow z^{\widetilde{K}_{nl}}(1\mp qz)^{\frac{1}{q}%
\left( \widetilde{S}_{l}+\frac{q}{2}\right) },\text{ }\widetilde{K}_{nl}>0,%
\text{ }\frac{1}{q}\left( \widetilde{S}_{l}+\frac{q}{2}\right) >0,
\end{equation}%
and, hence, the unnormalized wave functions are being expressed in terms of
the Jacobi polynomials as%
\begin{equation}
g^{\pm }(z)=\mathcal{N}_{nl}z^{\widetilde{K}_{nl}}(1\mp qz)^{\frac{1}{q}%
\widetilde{S}_{l}+\frac{1}{2}}P_{n}^{(2\widetilde{K}_{nl},\frac{2}{q}%
\widetilde{S}_{l})}(1-2qz),\text{ }z\in \lbrack 0,1/q]
\end{equation}%
where $\mathcal{N}_{nl}$ being the normalization constants and $P_{n}^{(2%
\widetilde{K}_{nl},\frac{2}{q}\widetilde{S_{l}})}(1-2qz)=\frac{\left( 2%
\widetilde{K}_{nl}+1\right) _{n}}{n!}_{2}F_{1}(-n,2\widetilde{K}_{nl}+\frac{2%
}{q}\widetilde{S_{l}}+n+1,2\widetilde{K}_{nl}+1;qz)$ with $(m)_{n}=\frac{%
\left( m+n-1\right) !}{\left( m-1\right) !}$ is Pochhammer's symbol$.$ For
example, if $q\geq 1$ then $z\in \lbrack 0,1/q]$ and if $q\leq -1$ then $%
z\in \lbrack 1/q,0]$ lie within or on the boundary of the interval $\left[
-1,+1\right] .$

Hence, the total wave function of the $q$-DEP/GDEP functions is 
\begin{equation*}
\psi _{\pm }(\overrightarrow{r})=\mathcal{N}_{nl}\frac{1}{r}\left[ \pm \exp
(-2\alpha r)\right] ^{\widetilde{K}_{nl}}\left[ 1-\pm q\exp (-2\alpha r)%
\right] ^{\frac{1}{q}\widetilde{S}_{l}+\frac{1}{2}}
\end{equation*}%
\begin{equation}
\times P_{n}^{(2\widetilde{K}_{nl},\frac{2}{q}\widetilde{S}_{l})}(1-\pm
2q\exp (-2\alpha r))Y_{lm}(\widehat{r}).
\end{equation}%
where the normalization constants $\mathcal{N}_{nl}$ are calculated
explicitly in Appendix B.

\section{Discussions}

In this section, we are going to study two special cases of the energy
eigenvalues given by Eq. (34). First, we consider the $s$-wave ($l=0$)
vibrational energy equation: 
\begin{equation*}
2\sqrt{\left( mc^{2}+E_{R}\right) D\left( 1-\sigma \right)
^{2}+m^{2}c^{4}-E_{R}^{2}}=
\end{equation*}%
\begin{equation}
\frac{4\left( mc^{2}+E_{R}\right) D\sigma -\left( \sqrt{4\left(
mc^{2}+E_{R}\right) D\sigma ^{2}+\alpha ^{2}\hbar ^{2}c^{2}}+\alpha \hbar
c\left( 2n+1\right) \right) ^{2}}{\sqrt{4\left( mc^{2}+E_{R}\right) D\sigma
^{2}+\alpha ^{2}\hbar ^{2}c^{2}}+\alpha \hbar c\left( 2n+1\right) },
\end{equation}%
where $n=0,1,2,\cdots ,n_{\max },$ where $n_{\max }$ is the number of bound
states for the whole bound spectrum near the continuous zone. $n_{\max }$ is
the largest integer which is less than or equal to the value of $n$ that
makes the right side of Eq. (39) vanish, that is, 
\begin{subequations}
\begin{equation}
n\rightarrow n_{\max }=\frac{1}{2}\left[ -1-\sqrt{\frac{4D}{\alpha ^{2}\hbar
^{2}c^{2}}\left( mc^{2}+E_{R}\right) \sigma ^{2}+1}+\sqrt{\frac{4D}{\alpha
^{2}\hbar ^{2}c^{2}}\left( mc^{2}+E_{R}\right) \sigma }\right] ,
\end{equation}%
\begin{equation}
E_{n_{\max }}^{(R)}\rightarrow mc^{2}+D_{e},
\end{equation}%
If $\sigma =1,$ then $E_{n_{\max }}^{(R)}\rightarrow mc^{2}.$ The
corresponding normalized wave functions can easily be found directly from
Eq. (38) as 
\end{subequations}
\begin{equation*}
R_{nl}^{(\pm )}(r)=\mathcal{N}_{n}\frac{1}{r}\left[ \pm \exp (-2\alpha r)%
\right] ^{\widetilde{k}_{n}}\left[ 1-\pm q\exp (-2\alpha r)\right] ^{\frac{1%
}{q}\widetilde{s}+\frac{1}{2}}
\end{equation*}%
\begin{equation}
\times P_{n}^{(2\widetilde{k}_{n},\frac{2}{q}\widetilde{s})}(1-\pm 2q\exp
(-2\alpha r)),
\end{equation}%
where $\mathcal{N}_{n}$ are the normalization constants and calculated in
Appendix B.

Second, we discuss the non-relativistic limit of the energy eigenvalues and
wave functions in the non-relativistic limit. Obviously, the currently
calculated KG solutions, under the previously mentioned transformations, can
be reduced to their associated Schr\"{o}dinger ones for the GDEP functions
as 
\begin{equation*}
E_{nl}^{(R)}\rightarrow E_{nl}^{(NR)}=D_{e}+\frac{l(l+1)\hbar ^{2}}{2\mu
r_{e}^{2}}A_{0}
\end{equation*}%
\begin{equation}
-\frac{\alpha ^{2}\hbar ^{2}}{2\mu }\left[ \frac{\left( \frac{S_{l}}{q}%
\right) ^{2}-\frac{Q_{l}}{q}-\frac{1}{4}-\left( \frac{S_{l}}{q}+n+\frac{1}{2}%
\right) ^{2}}{\frac{S_{l}}{q}+n+\frac{1}{2}}\right] ^{2},\text{ }%
n,l=0,1,2,\cdots ,
\end{equation}%
or more explicitly as%
\begin{equation*}
E_{nl}^{(NR)}=D_{e}+\frac{l(l+1)\hbar ^{2}}{2\mu r_{e}^{2}}A_{0}
\end{equation*}%
\begin{equation}
-\frac{\alpha ^{2}\hbar ^{2}}{2\mu }\left[ \frac{\frac{2\mu D}{\hbar
^{2}\alpha ^{2}}\sigma +\frac{l(l+1)}{4\alpha ^{2}r_{e}^{2}}\left( \frac{%
A_{2}}{q^{2}}-\frac{A_{1}}{q}\right) -\left( n+\frac{1}{2}+\sqrt{\frac{2\mu D%
}{\hbar ^{2}\alpha ^{2}}\sigma ^{2}+\frac{l(l+1)}{4\alpha ^{2}r_{e}^{2}}%
\frac{A_{2}}{q^{2}}+\frac{1}{4}}\right) ^{2}}{n+\frac{1}{2}+\sqrt{\frac{2\mu
D}{\hbar ^{2}\alpha ^{2}}\sigma ^{2}+\frac{l(l+1)}{4\alpha ^{2}r_{e}^{2}}%
\frac{A_{2}}{q}+\frac{1}{4}}}\right] ^{2},
\end{equation}%
which is identical to Eq. (28) of Ref. [33] if one sets $\delta =1$ and $%
q=1. $ This represents the approximate Schr\"{o}dinger solution of Eq. (2)
for the ro-vibratinal molecules. The non-relativistic limits for the
vibrational energy states ($l=0$) read

\begin{equation}
E_{n}^{(NR)}=D_{e}-\frac{\alpha ^{2}\hbar ^{2}}{2\mu }\left[ \frac{\frac{%
2\mu D}{\hbar ^{2}\alpha ^{2}}\sigma -\left( n+\frac{1}{2}+\sqrt{\frac{2\mu D%
}{\hbar ^{2}\alpha ^{2}}\sigma ^{2}+\frac{1}{4}}\right) ^{2}}{n+\frac{1}{2}+%
\sqrt{\frac{2\mu D}{\hbar ^{2}\alpha ^{2}}\sigma ^{2}+\frac{1}{4}}}\right]
^{2},\text{ }n=0,1,2,\cdots ,n_{\max },
\end{equation}%
and the corresponding unnormalized wave functions from Eq. (38) are 
\begin{equation*}
\psi _{\pm }(\overrightarrow{r})=\mathcal{N}_{nl}\frac{1}{r}\left[ \pm \exp
(-2\alpha r)\right] ^{K_{nl}}\left[ 1-\pm q\exp (-2\alpha r)\right] ^{\frac{1%
}{q}S_{l}+\frac{1}{2}}
\end{equation*}%
\begin{equation}
\times P_{n}^{(2K_{nl},\frac{2}{q}S_{l})}(1-\pm 2q\exp (-2\alpha r))Y_{lm}(%
\widehat{r}).
\end{equation}%
where $K_{nl}$ and $S_{l}$ are defined in Eq. (27) and the condition for $%
n_{\max }$ turns to become%
\begin{equation}
n\rightarrow n_{\max }=\frac{1}{2}\left[ -1-\sqrt{\frac{8\mu D}{\hbar
^{2}\alpha ^{2}}\sigma ^{2}+1}+\sqrt{\frac{8\mu D}{\hbar ^{2}\alpha ^{2}}%
\sigma }\right] ,\text{ }E_{n_{\max }}^{(NR)}\rightarrow D_{e}.
\end{equation}%
Thus, the finiteness of $n_{\max }$ is reflected in the above condition. If $%
\sigma =1,$ then $n_{\max }\rightarrow 0.$

\section{Applications to Diatomic Molecules}

We have calculated the non-relativistic energy states for the two selected $%
H_{2}$ and $Ar_{2}$ diatomic molecules using energy equation (35) with $%
q\rightarrow 1$ and (23). The spectroscopic constants of these two molecules
are given in Table 1. The vibrating ground state energy eigenvalues $%
E_{+}^{0}$ (in $cm^{-1}$) for the $H_{2}$ molecule in the non-deformed EP
functions $V_{+}(r)$ are found using the NU method for the potential
parameters given in Table 2. Our numerical results obtained in the present
NU model are listed together with the numerical results obtained by using SC
(as Semi-Classical) procedure and a QM (as Quantum-Mechanical) method
mentioned in Ref. [32] for various potential parameters. Obviously, as shown
in Table 2, the results obtained in the present model are in high agreement
with those obtained by QM. However, the SC procedure is proportionally
different. Therefore, the differences between our results and SC procedure
are less than $0.01$ $cm^{-1},$ i.e., they are negligible because of these
approximations: $1$ $a.m.u=931.502$ $MeV/c^{2},$ $1$ $cm^{-1}=1.23985\times
10^{-4}$ $eV$ and $\hbar c=1973.29$ $eV.A^{\circ }$ [57]. The second
application is applied to $Ar_{2}$ molecule. We confine our study to
calculate the ro-vibrating energy states for the $V_{+}(r)$ potential using
the following potential parameters: $\sigma =25.23,$ $\delta =41.75$ and $%
\alpha =0.6604$ $(A^{\circ })^{-1}$ [32] together with the parameters given
in Table 1. For the previously given set of physical parameters, we plot the
non-relativistic energy spectrum curve as a function of vibrational quantum
number $n$ as seen in Figure 1$.$ Obviously, in a reference to Figure 1, the
energy spectrum of the diatomic molecule $Ar_{2}$ approaches the value of $%
D_{e}$ as $n$ approaches $n_{\max }=6.689$ or $n_{\max }=6.$ This is also
verified analytically from Eqs. (44) and (46). Moreover, a plot of the
non-relativistic energy spectrum curve as a function of the potential
strength $D_{e}$ for the above given set of physical parameters and $n=0$
for $Ar_{2}$ molecule is shown in Figure 2. The relationship is noticed to
be nearly linear for any arbitrary value of vibrational quantum number $n$.
The attractive energy value increases with the increasing potential
strength. The splittings of the energy states of $s$-waves $%
E_{+}=E_{+}(n\neq 0)-E_{+}(n=0)$ obtained by the NU method and SC procedures
are presented in Table 3. The present results $\Delta E_{+}(NU)$ from NU
method and $\Delta E_{+}(SC)$ obtained from the SC procedures are also
compared with four-different experimental results labeled $\Delta
E(a),\Delta E(b),$ $\Delta E(c)$ and $\Delta E(d)$ taken from Ref. [32]$.$
It is obvious from Table 3 that our results are very close with the
experimentally determined values as well as the SC procedure results.
Finally, the approximated rotating and vibrating energy states of the $%
V_{+}(r)$ given in Eq. (2) for the $Ar_{2}$ and $H_{2}$ molecules are also
calculated for the $l\neq 0$ case. Table 4 shows the energy levels for
vibrational $(n=0,1,2,3,4,5)$ and rotational $(l=0,1,2)$ quantum numbers.

\section{Conclusions}

To summarize, we have presented the approximate bound state energy\
eigenvalues and their corresponding normalized wave functions of the
relativistic spin-$0$ particle in the radial ($3+1$)-dimensional KG equation
with equal scalar and vector $q$-DEP/GDEP functions by means of the
parametric generalization of the NU method. We point out that the KG wave
functions are found in terms of the Jacobi polynomials. The analytic
expressions for the relativistic energy expression and the corresponding
wave functions of this molecular system can be reduced to the well-known
non-relativistic solutions and to the $s$-waves solutions as well. The
relativistic energy $E_{R}$ defined implicitly by Eq. (33) is rather a
transcendental equation and it has many solutions for any arbitrarily chosen
values of usual quantum numbers $n$ and $l.$ The method presented in this
paper is general and worth extending to the solution of other molecular
interaction problems. The method is simple and useful in solving other
complicated systems analytically without giving any restriction on the
solution of some quantum systems as is the case in the other models. We have
also seen that for the nonrelativistic model, the approximate energy
spectrum can be obtained either by directly solving the Schr\"{o}dinger
equation [41] or rather by even applying appropriate transformations to the
relativistic solution as currently shown. We should emphasize that the
approximate bound state energy spectrum obtained in the present work might
have some interesting applications in different branches like atomic and
molecular physics and quantum chemistry. The present solution is describing
the inter-molecular structures and interactions in diatomic molecules
[32-36,41,58,59]. The present study is also useful in calculating the
vibrating energy for different radial $n$ quantum numbers as well as the
rotating energy for different orbital $l$ quantum numbers. To conclude, the
proposed $q$-deformation potential with a flexible and fixed value $q$ (real
or complex) can generate various potential models with various energy
solutions$.$

\acknowledgments The author thanks the three anonymous kind referees for the
very constructive comments and suggestions. He is also grateful for the
partial support provided by the Scientific and Technological Research
Council of Turkey (T\"{U}B\.{I}TAK).\bigskip

\appendix

\section{Parameterized Version of the NU Method}

We complement the theoretical formulation of the NU method in presenting the
essential polynomials, energy equation and wave functions together with
their relevant constants as follows.

(i) The key polynomials:%
\begin{equation}
\pi (z)=c_{4}+c_{5}z-\left[ \left( \sqrt{c_{9}}+c_{3}\sqrt{c_{8}}\right) z-%
\sqrt{c_{8}}\right] ,
\end{equation}%
\begin{equation}
k=-\left( c_{7}+2c_{3}c_{8}\right) -2\sqrt{c_{8}c_{9}}.
\end{equation}%
\begin{equation}
\tau (z)=1-\left( c_{2}-2c_{5}\right) z-2\left[ \left( \sqrt{c_{9}}+c_{3}%
\sqrt{c_{8}}\right) z-\sqrt{c_{8}}\right] ,
\end{equation}%
\begin{equation}
\tau ^{\prime }(z)=-2c_{3}-2\left( \sqrt{c_{9}}+c_{3}\sqrt{c_{8}}\right) <0,
\end{equation}%
(ii) The energy equation:%
\begin{equation}
\left( c_{2}-c_{3}\right) n+c_{3}n^{2}-\left( 2n+1\right) c_{5}+\left(
2n+1\right) \left( \sqrt{c_{9}}+c_{3}\sqrt{c_{8}}\right) +c_{7}+2c_{3}c_{8}+2%
\sqrt{c_{8}c_{9}}=0.
\end{equation}%
(iii) The wave functions:%
\begin{equation}
\rho (z)=z^{c_{10}}(1-c_{3}z)^{c_{11}},
\end{equation}%
\begin{equation}
\phi (z)=z^{c_{12}}(1-c_{3}z)^{c_{13}},\text{ }c_{12}>0,\text{ }c_{13}>0,
\end{equation}%
\begin{equation}
y_{n}(z)=P_{n}^{\left( c_{10},c_{11}\right) }(1-2c_{3}z),\text{ }c_{10}>-1,%
\text{ }c_{11}>-1,\text{ }z\in \left[ 0,1/c_{3}\right] ,
\end{equation}%
\begin{equation}
u(z)=\mathcal{N}_{n}z^{c_{12}}(1-c_{3}z)^{c_{13}}P_{n}^{\left(
c_{10},c_{11}\right) }(1-2c_{3}z),
\end{equation}%
where the Jacobi polynomial $P_{n}^{\left( \mu ,\nu \right) }(x)$ is defined
only for $\mu >-1,$ $\nu >-1,$ and for the argument $x\in \left[ -1,+1\right]
$ and $\mathcal{N}_{n}$ is a normalizing factor.$.$ It can be expressed in
terms of the hypergeometric function as%
\begin{equation}
P_{n}^{\left( \mu ,\nu \right) }(1-2s)=\frac{\left( \mu +1\right) _{n}}{n!}%
\begin{array}{c}
_{2}F_{1}%
\end{array}%
\left( -n,1+\mu +\nu +n;\mu +1;s\right) ,
\end{equation}%
where $s\in \left[ 0,1\right] $ which lie within or on the boundary of the
interval $\left[ -1,1\right] .$ Also, the above wavefunctions can be
expressed in terms of the hypergeometric function as%
\begin{equation}
u(z)=\mathcal{N}_{n}z^{c_{12}}(1-c_{3}z)^{c_{13}}%
\begin{array}{c}
_{2}F_{1}%
\end{array}%
\left( -n,1+c_{10}+c_{11}+n;c_{10}+1;c_{3}z\right) ,
\end{equation}%
where $c_{12}>0,$ $c_{13}>0$ and $z\in \left[ 0,1/c_{3}\right] .$

(iv) The relevant constants:%
\begin{equation*}
c_{4}=\frac{1}{2}\left( 1-c_{1}\right) ,\text{ }c_{5}=\frac{1}{2}\left(
c_{2}-2c_{3}\right) ,\text{ }c_{6}=c_{5}^{2}+B_{1},
\end{equation*}%
\begin{equation*}
\text{ }c_{7}=2c_{4}c_{5}-B_{2},\text{ }c_{8}=c_{4}^{2}+B_{3},\text{ }%
c_{9}=c_{3}\left( c_{7}+c_{3}c_{8}\right) +c_{6},
\end{equation*}%
\begin{equation*}
c_{10}=c_{1}+2c_{4}+2\sqrt{c_{8}}-1>-1,\text{ }c_{11}=1-c_{1}-2c_{4}+\frac{2%
}{c_{3}}\sqrt{c_{9}}>-1,
\end{equation*}%
\begin{equation}
c_{12}=c_{4}+\sqrt{c_{8}}>0,\text{ }c_{13}=-c_{4}+\frac{1}{c_{3}}\left( 
\sqrt{c_{9}}-c_{5}\right) >0.
\end{equation}

$\label{appendix}$

\section{Normalization of the radial wave function}

In order to find the normalization factor $\mathcal{N}_{nl}$, we start by
writting the normalization condition:%
\begin{equation}
\frac{\mathcal{N}_{nl}^{2}}{2\alpha }\int_{0}^{1}z^{2\widetilde{K}%
_{nl}-1}(1-z)^{2\widetilde{S}_{l}+1}\left[ P_{n}^{(2\widetilde{K}_{nl},2%
\widetilde{S}_{l})}(1-2z)\right] ^{2}dz=1,
\end{equation}%
where $q=1.$ Unfortunately, there is no formula available to calculate this
key integration. Neveretheless, we can find the explicit normalization
constant $\mathcal{N}_{nl}.$ For this purpose, it is not difficult to obtain
the results of the above integral by using the following formulas [59]%
\begin{equation}
\dint\limits_{0}^{1}\left( 1-s\right) ^{\mu -1}s^{\nu -1}%
\begin{array}{c}
_{2}F_{1}%
\end{array}%
\left( \alpha ,\beta ;\gamma ;as\right) dz=\frac{\Gamma (\mu )\Gamma (\nu )}{%
\Gamma (\mu +\nu )}%
\begin{array}{c}
_{3}F_{2}%
\end{array}%
\left( \nu ,\alpha ,\beta ;\mu +\nu ;\gamma ;a\right) ,
\end{equation}%
and $%
\begin{array}{c}
_{2}F_{1}%
\end{array}%
\left( a,b;c;z\right) =\frac{\Gamma (c)}{\Gamma (a)\Gamma (b)}%
\dsum\limits_{p=0}^{\infty }\frac{\Gamma (a+p)\Gamma (b+p)}{\Gamma (c+p)}%
\frac{z^{p}}{p!}.$ Following Ref. [59], we calculate the normalization
constants:%
\begin{equation}
\mathcal{N}_{nl}=\left[ \frac{\Gamma (2\widetilde{K}_{nl}+1)\Gamma (2%
\widetilde{S}_{l}+2)}{2\alpha \Gamma (n)}\dsum\limits_{m=0}^{\infty }\frac{%
(-1)^{m}\left( 1+n+2(\widetilde{K}_{nl}+\widetilde{S}_{l})\right) _{m}\Gamma
(n+m)}{m!\left( m+2\widetilde{K}_{nl}\right) !\Gamma \left( m+2(\widetilde{K}%
_{nl}+\widetilde{S}_{l}+1)\right) }f_{nl}\right] ^{-1/2}\text{ ,}
\end{equation}%
where 
\begin{equation}
f_{nl}=%
\begin{array}{c}
_{3}F_{2}%
\end{array}%
\left( 2\widetilde{K}_{nl}+m,-n,n+1+2(\widetilde{K}_{nl}+\widetilde{S}%
_{l});m+2(\widetilde{K}_{nl}+\widetilde{S}_{l}+1);1+2\widetilde{K}%
_{nl};1\right) .
\end{equation}%
Furthermore, the normalization constants for the $s$-wave can be also found
as%
\begin{equation}
\mathcal{N}_{n}=\left[ \frac{\Gamma (2\widetilde{k}_{n}+1)\Gamma (2%
\widetilde{s}+2)}{2\alpha \Gamma (n)}\dsum\limits_{m=0}^{\infty }\frac{%
(-1)^{m}\left( 1+n+2(\widetilde{k}_{n}+\widetilde{s})\right) _{m}\Gamma (n+m)%
}{m!\left( m+2\widetilde{k}_{n}\right) !\Gamma \left( m+2(\widetilde{k}_{n}+%
\widetilde{s}+1)\right) }g_{n}\right] ^{-1/2}\text{ ,}
\end{equation}%
where 
\begin{equation}
g_{n}=%
\begin{array}{c}
_{3}F_{2}%
\end{array}%
\left( 2\widetilde{k}_{n}+m,-n,n+1+2(\widetilde{k}_{n}+\widetilde{s});m+2(%
\widetilde{k}_{n}+\widetilde{s}+1);1+2\widetilde{k}_{n};1\right) ,
\end{equation}%
and 
\begin{equation*}
\widetilde{k}_{n}=\frac{1}{2\alpha \hbar }\sqrt{\left( mc^{2}+E_{R}\right)
D\left( 1-\sigma \right) ^{2}+m^{2}c^{4}-E_{R}^{2}},
\end{equation*}%
\begin{equation}
\widetilde{s}=\frac{q}{2\alpha \hbar }\sqrt{4\left( mc^{2}+E_{R}\right)
D\sigma ^{2}+\alpha ^{2}\hbar ^{2}},\text{ }n=0,1,2,\cdots ,
\end{equation}%
where $E_{R}$ is the solution of the transcendental equation (33).

\ {\normalsize 
}\FRAME{ftbpFO}{0.0277in}{0.0277in}{0pt}{\Qct{A plot of the non-relativistic
energy spectrum curve as a function of the vibrational quantum number $n$
for a given set of physical parameters for $Ar_{2}$ molecule.}}{\Qlb{1}}{%
Figure}{}

\FRAME{ftbpFO}{0.0277in}{0.0277in}{0pt}{\Qct{A plot of the non-relativistic
energy spectrum curve as a function of the potential strength $D_{e}$ for a
given set of physical parameters and vibrational ground state $n=0$ for $%
Ar_{2}$ molecule. }}{}{Figure}{}\ {\normalsize 
}

\baselineskip= 2\baselineskip
\ 
\begin{table}[tbp]
\caption{ The spectroscopic constants of the EP for $H_{2}$ and $Ar_{2}$
molecules [26].}%
\begin{tabular}{lll}
\tableline Parameters & $H_{2}$ & $Ar_{2}$ \\ 
\tableline\tableline$D_{e}$ $(cm^{-1})$ & 38281 & 99.55 \\ 
$r_{e}$ $(A^{\circ })$ & 0.7414 & 3.759 \\ 
$\mu $ (a.m.u) & 0.50407 & 19.9812 \\ 
\tableline &  & 
\end{tabular}%
\end{table}
\bigskip \bigskip

\begin{table}[tbp]
\caption{The EP parameters of the $V_{+}(r)$ and the ground state energy, $%
E_{+}^{00}$ (in $cm^{-1}$) of the $H_{2}$ molecule.}%
\begin{tabular}{llllll}
\tableline$\sigma $ & $\delta $ & $\alpha $ $(A^{\circ })^{-1}$ & $E_{+}(SC)$
& $E_{+}(QM)$ & Present \\ 
\tableline\tableline$426.826$ & $463.102$ & $0.9327$ & $2167.68$ & $2168.93$
& 2168.68 \\ 
$47.294$ & $102.341$ & $0.6146$ & $2153.69$ & $2164.83$ & 2164.45 \\ 
$28.685$ & $117.121$ & $0.3826$ & $2139.57$ & $2157.69$ & 2157.53 \\ 
$21.250$ & $213.212$ & $0.1762$ & $2124.29$ & $2148.40$ & 2147.53 \\ 
\tableline &  &  &  &  & 
\end{tabular}%
\end{table}
\begin{table}[tbp]
\caption{Comparisons of experimentally calculated $s$-states energy
transition values $\Delta E_{n,0}(cm^{-1})$ for $n\neq 0\rightarrow n=0$
together with the results of the SC procedure and the present NU method for
the $Ar_{2}$ molecule.}%
\begin{tabular}{lllllll}
\tableline$n$ & Present & $\Delta E(a)$ & $\Delta E(b)$ & $\Delta E(c)$ & $%
\Delta E(d)$ & $\Delta E_{+}(SC)$ \\ 
\tableline\tableline1 & 25.808 & 25.74 & 25.49 & 25.21 & 25.56 & 25.75 \\ 
2 & 46.079 & 46.15 & 45.63 & 45.02 & 46.00 & 46.01 \\ 
3 & 61.472 & 61.75 & 60.70 & 60.04 & 61.32 & 61.42 \\ 
4 & 72.536 & 72.66 & 71.33 & 70.92 & 71.52 & 72.52 \\ 
5 & 79.733 & 79.44 & - & - & - & 79.79 \\ 
6 & 83.453 & - & - & - & - & 83.59 \\ 
7 & 84.026 & - & - & - & - & - \\ 
\tableline &  &  &  &  &  & 
\end{tabular}%
\end{table}

\begin{table}[tbp]
\caption{Energy levels $E_{n,l}(cm^{-1})$ for $Ar_{2}$ and $H_{2}$ molecules
in $V_{+}(r)$ u$\sin $g the NU method.}%
\begin{tabular}{llll}
\tableline$n$ & $l$ & $E_{+}(Ar_{2})$ & $E_{+}(H_{2})$ \\ 
\tableline\tableline0 & 0 & 15.3828 & 2168.68 \\ 
1 & 0 & 41.1910 & 6306.66 \\ 
& 1 & 25.7584 & 6331.10 \\ 
2 & 0 & 61.4619 & 10183.8 \\ 
& 1 & 49.7874 & 10207.6 \\ 
& 2 & - & 10255.2 \\ 
3 & 0 & 76.8546 & 13802.1 \\ 
& 1 & 68.3028 & 13825.2 \\ 
& 2 & 19.9133 & 13871.5 \\ 
4 & 0 & 87.9188 & 17163.2 \\ 
& 1 & 82.0041 & 17185.7 \\ 
& 2 & 46.4777 & 17230.7 \\ 
5 & 0 & 95.1159 & 20269.1 \\ 
& 1 & 91.4672 & 20291.0 \\ 
& 2 & 66.5474 & 20334.8 \\ 
\tableline &  &  & 
\end{tabular}%
\end{table}

\bigskip \bigskip

\end{document}